\documentclass[11pt]{article}

\usepackage{array,fullpage,multirow}
\usepackage{amssymb,amsmath,amsthm,sectsty,url,nicefrac,color}
\usepackage[bookmarks=true,pdfstartview=FitH,colorlinks,linkcolor=blue,filecolor=blue,citecolor=blue,urlcolor=blue]{hyperref}
\usepackage{cleveref}
\usepackage{aliascnt}
\usepackage[numbers]{natbib} 
\usepackage[inline]{enumitem}
\usepackage{subcaption}
\usepackage{graphicx}

\usepackage[boxed]{algorithm}
\usepackage{algpseudocode}

\numberwithin{equation}{section}

\theoremstyle{nicetheorem}

\newtheorem{theorem}{Theorem}[section]

\newaliascnt{lemma}{theorem}
\newtheorem{lemma}[lemma]{Lemma}
\aliascntresetthe{lemma}
\crefname{lemma}{Lemma}{Lemmas}

\newaliascnt{example}{theorem}

\aliascntresetthe{example}
\crefname{example}{Example}{Examples}

\newaliascnt{claim}{theorem}
\newtheorem{claim}[claim]{Claim}
\aliascntresetthe{claim}
\crefname{claim}{Claim}{Claims}

\newaliascnt{corollary}{theorem}

\aliascntresetthe{corollary}
\crefname{corollary}{Corollary}{Corollaries}

\newaliascnt{construction}{theorem}

\aliascntresetthe{construction}
\crefname{construction}{Construction}{Constructions}

\newaliascnt{question}{theorem}
\newtheorem{question}[question]{Question}
\aliascntresetthe{question}
\crefname{question}{Question}{Questions}

\newaliascnt{fact}{theorem}

\aliascntresetthe{fact}
\crefname{fact}{Fact}{Facts}

\newaliascnt{proposition}{theorem}

\aliascntresetthe{proposition}
\crefname{proposition}{Proposition}{Propositions}

\newaliascnt{conjecture}{theorem}

\aliascntresetthe{conjecture}
\crefname{conjecture}{Conjecture}{Conjectures}

\newaliascnt{definition}{theorem}
\newtheorem{definition}[definition]{Definition}
\aliascntresetthe{definition}
\crefname{definition}{Definition}{Definitions}

\newaliascnt{remark}{theorem}
\newtheorem{remark}[remark]{Remark}
\aliascntresetthe{remark}
\crefname{remark}{Remark}{Remarks}

\newaliascnt{observation}{theorem}
\newtheorem{observation}[observation]{Observation}
\aliascntresetthe{observation}
\crefname{observation}{Observation}{Observations}

\crefname{algorithm}{Algorithm}{Algorithms}

\newaliascnt{notation}{theorem}
\newtheorem{notation}[notation]{Notation}
\aliascntresetthe{notation}
\crefname{notation}{Notation}{Notations}

\newcommand{\tri}{$\exists \forall$-Triangle Problem}
\newcommand{\tris}{$\exists \forall$-Triangle Problem }

\newcommand{\beq}{\begin{equation}}
\newcommand{\eeq}{\end{equation}}

\newcommand{\kout}{k_\text{out}}

\usepackage{tikz}

\newcommand{\remove}[1]{}
\definecolor{green1}{rgb}{0.40, 0.8, 0.2}

\def\({\left(}
\def\){\right)}
\makeindex

\title{From Donkeys to Kings in Tournaments}

\author{
Amir Abboud\thanks{Department of Computer Science and Applied Mathematics,   Weizmann Institute of Science,  Rehovot 76100, Israel. Email: \texttt{amir.abboud@weizmann.ac.il}. This work is part of the project CONJEXITY, which has received funding from the European Research Council (ERC) under the European Union's Horizon Europe research and innovation program (grant agreement No.~101078482). Supported by an Alon scholarship and a research grant from the Center for New Scientists at the Weizmann Institute of Science.}
\and
Tomer Grossman\thanks{Department of Computer Science and Applied Mathematics,
   Weizmann Institute of Science,  Rehovot 76100, Israel. Email: \texttt{tomer.grossman@weizmann.ac.il}.}
\and
 Moni Naor\thanks{Department of Computer Science and Applied Mathematics,
   Weizmann Institute of Science,  Rehovot 76100, Israel. Email:
   \texttt{moni.naor@weizmann.ac.il}. Supported in part by grants from the Israel Science Foundation (no.2686/20), by the Simons Foundation Collaboration on the Theory of Algorithmic Fairness and by the Israeli Council for Higher
Education (CHE) via the Weizmann Data Science Research Center. Incumbent of the Judith Kleeman Professorial
   Chair.}
   \and
Tomer Solomon\thanks{Tel Aviv University.
Email: \texttt{tomer.solomon1@gmail.com}. This material is based upon work supported by DARPA under Agreement
  No. HR00110C0086.  Any opinions, findings and conclusions or
  recommendations expressed in this material are those of the author(s)
  and do not necessarily reflect the views of the United States
  Government or DARPA.}
}
 \date{}

\begin{document}

\maketitle

\begin{abstract}
A tournament is an orientation of a complete graph. A vertex that can reach every other vertex within two steps is called a \emph{king}. We study the complexity of finding $k$ kings in a tournament graph. 

We show that the randomized query complexity of finding $k \le 3$ kings is $O(n)$, and for the deterministic case it takes the same amount of queries (up to a constant) as finding a single king (the best known deterministic algorithm makes $O(n^{3/2})$ queries). On the other hand, we show that finding $k \ge 4$ kings requires  
$\Omega(n^2)$ queries, even in the randomized case.

We consider the RAM model for $k \geq 4$. We show an algorithm that finds $k$ kings in time $O(kn^2)$, which is optimal for constant values of $k$. Alternatively, one can also find $k \ge 4$ kings in time $n^{\omega}$ (the time for matrix multiplication). We provide evidence that this is optimal for large $k$ by suggesting a fine-grained reduction from a variant of the triangle detection problem.

\end{abstract}
\newpage \setcounter{page}{1}
\section{Introduction}
A tournament is an orientation of a complete graph on $n$ vertices; that is, between any two distinct vertices $u,v$, there is exactly one edge, either $u \to v$ or $v \mapsto u$ (but not both).  
Often we think of this as a tournament where each vertex is a player and all players play against all other players, where $u$ beats $v$ then there is an edge from $u$ to $v$. Tournaments are a very basic combinatorial structure. A fundamental object we can study in tournaments is their kings:

\begin{definition}[King]
We say that a vertex $v$ of a tournament graph is a king if for every other vertex $u$ there is a path from $v$ to $u$ of length at most 2. 
\end{definition}

It is well known that every tournament contains at least one king (for instance, a vertex with the  {\em maximum} outdegree is necessarily a king). Some tournaments contain just one king (e.g.\ the transitive tournament\footnote{A transitive tournament is a tournament with the property that for all vertices $a,b,c$ if $a \to b$ and $b \to c$ then $a \to c$.}), whereas others contain many kings (e.g.\ in a {\em random} tournament {\em all} the nodes are kings with high probability). 
The complexity of finding kings is interesting because it is arguably the simplest combinatorial structure of tournament graphs, yet, the complexity of finding kings is not well understood.

\paragraph*{Models:} We are interested in the query model, that is where the important measure is the number of probes to the adjacency matrix, when the complexity is sublinear (i.e.\ $o(n^2)$, since the input size is ${n \choose 2} $), as well as in the standard RAM model in the regime when we cannot achieve sublinearity. In this regime every algorithm has to query the entire input and the question is how much computation is needed.  The question of the query complexity of finding one king has received attention in the query-based model~\cite{ShenSW03, AjtaiFHN16,BiswasJRS22, MandePS23}. We concentrate on the case of finding {\em multiple} kings:

\begin{center}
\textit{ What is the complexity of finding $k$ kings in a tournament graph?
}
\end{center}

Interestingly we discovered that there is a ``phase transition": when $k \le 3$ finding $k$ kings is a relatively easy problem and can be done in sublinear time, even by deterministic algorithms. On the other hand, when $k \ge 4$, even randomized algorithms require $\Omega(n^2)$ time. 

\subsection{Our Results}
We characterize the complexity of finding $k$ kings by providing both upper and lower bounds as a function of $k$. 

We characterize the complexity of finding kings in the query model, i.e.\ when we do not take into account any computation except edge queries. We note that our upper bounds can actually be implemented efficiently, that is we do not abuse this freedom. 

We start by providing a simple randomized time linear algorithm for finding a king, improving the result of \cite{MandePS23} who showed a $n \log \log n$ algorithm. While we did not find in the literature any explicit claim for a linear time randomized algorithm, the proposed one is very similar to \cite{ShenSW03} who show that finding the \emph{sorted} sequence of kings can be done roughly by a randomized quick sort with $n \log n$ queries. 
\begin{theorem} \label{thm:singlekingrand}
There exists a randomized algorithm for finding a king that makes $2n$ queries in expectation.
\end{theorem}

The best-known deterministic algorithm for finding a single king takes $O(n^{3/2})$ queries \cite{ShenSW03}. We generalize the results of finding a single king to finding 2 or 3 kings for both deterministic and randomized algorithms: 
\begin{theorem} \label{thm:multiplekings}
There exists a randomized algorithm that finds 2 or 3 kings in time $O(n)$ and there exists a deterministic algorithm that finds 2 or 3 kings in time $O(n^{3/2})$.

\end{theorem}

On the other hand, finding 4 or more kings is hard in the query model:
\begin{theorem} \label{thm:ManyKingsLower}
Finding 4 or more kings requires $\Omega(n^2)$ queries, even for randomized algorithms. 
\end{theorem}
The proof of this theorem uses the fact that it’s hard to distinguish kings from donkeys (i.e. nodes that would be kings if a single edge is flipped).

Since finding $k \geq 4$ kings requires $\Omega(n^2)$ queries in the query model, and thus the trivial algorithm of querying all the edges is optimal in this respect. For studying the complexity of finding $k \geq 4$ kings
we turn to the standard Word-RAM model in order to take into account the actual computation. Here, for small $k$ we suggest an $O(k n^2)$ algorithm and when $k$ is large or finding all kings we suggest an $O(n^{\omega})$ algorithm 
(where $n^{\omega}$ is the time for $n \times n$ matrix multiplication). Currently, the best-known value for $\omega$ is $\omega \approx 2.371552$ \cite{williamsXXZ}.  

\begin{theorem} \label{thm:ManyKingsUpper}
There exists a deterministic algorithm that finds $k$ kings in time complexity $O(\min\{n^{\omega}, kn^2\})$.
\end{theorem}

Finally, we show that the above is likely optimal by showing a reduction from the \tri: Given a tripartite graph on sets $A,B,C$ deciding if every edge between $A$ and $C$ belongs to some triangle. See \cref{ref:TriDetBackground} for more detail about the \tri.

\begin{theorem} [Informal] \label{thm:finegrained}
Assuming that the \tris cannot be solved faster than $n^{\omega}$, then $n^{\omega}$ is the optimal algorithm for finding $k$ kings when $k \in \Omega(n)$. 
\end{theorem}

\subsection{Related Work}


Interest in the notion of Kings started in the mathematical biology literature, in particular the work of Hyman Landau~\cite{Landau53}. 
See Maurer~\cite{Maurer} for many basic mathematical properties about kings. Shen, Sheng and Wu~\cite{ShenSW03} showed that a king can be found in $O(n^{3/2})$ and gave a lower bound of $\Omega(n^{4/3})$. Ajtai et al.~\cite{AjtaiFHN16} generalized the above results for finding a vertex that is within distance $d$ to every other vertex.  Their setting is for imprecise comparisons when it is hard to distinguish similar physical stimuli, e.g. by human subjects, and the goal is to minimize the number of comparisons for tasks such a max finding, selection and sorting. 
Biswas et al.~\cite{BiswasJRS22} study the $d$-king cover problem -- that is finding a set, $S$ where every vertex is reachable within distance $d$ from some vertex in $S$. 
Lastly, they study the streaming problems of determining whether a new vertex added to a tournament is a king. Mande, Paraashar and Saurabh~\cite{MandePS23} gave an ${O}(n \log\log n)$ randomized algorithm and an $\tilde{O}(\sqrt{n})$ quantum algorithm for finding a king; these bounds are tight up to logarithmic factors.

The query complexity of other problems on tournaments was studied as well. For instance, Hamiltonian paths via comparison based sorting, resulting with complexity $O(n \log n)$~\cite{DBLP:journals/siamdm/Bar-NoyN90}. Other types of problems are those related to voting notions. Finding whether there is a `Condorcet winner', i.e.\ a node of degree $n-1$, takes 
$2n-\lfloor \log n \rfloor -2$
queries~\cite{BalasubramanianRS97,Procaccia08}. Other problems in voting, such as finding the Copeland set, require $\Omega(n^2)$ queries~\cite{MaitiD24}. The problem of finding the maximum degree node requires $\Omega(n^2)$ queries and is close but not quite evasive~\cite{GoyalJR20}.

Lastly, one of the first fine-grained reductions is related to tournaments, by Megiddo and Vishkin~\cite{MegiddoV88} showing that the problem of finding a minimum dominating set in a tournament (we know that there exists one of size at most $\log n$) is as hard as solving satisfiability with $O(\log^2 n)$ variables. In modern terminology, they showed ETH hardness.

\subsection{Open Problems}
\begin{question}
What is the deterministic query complexity of finding a single king?

The bounds suggested in Shen, Sheng and Wu~\cite{ShenSW03} are still the best - somewhere between the $\Omega(n^{4/3})$ lower bound and the $O(n^{3/2})$ upper bound 
\end{question}

We show that the complexity of finding $k \in \Omega(n)$ kings is at least as hard as a variant of the triangle detection problem, and at least as easy as matrix multiplication. 

\begin{question}
Is there a fine grained reduction from the $k$-king problem to \tri?
\end{question}

We introduce a variant of the triangle detection problem, where the quantifiers are changed. It is interesting to understand the relationship between these two problems.
\begin{question}
What is the relationship between \tris and the more common problem of triangle detection. Are they subcubic equivalent? 
\end{question}
\section{Preliminaries}

A tournament graph is a complete directed graph. A king in a tournament graph is a vertex that is within distance two of every other vertex. We assume that the tournament is given via an adjacency matrix, and each query inputs two vertices, $u$ and $v$ and returns the orientation of the edge between $u$ and $v$. More formally:
 
\begin{definition}
A graph $G = (V,E)$ is defined by two sets, $V,E$ where $V$ is the set of \emph{Vertices} and $E$ is the set of \emph{Edges}.
We call the graph \emph{undirected} if an edge $e \in E$ is a (unordered) set of two vertices, i.e. $e = \{v,u\}$ for $v,u \in V$. Similarly, we say a graph is \emph{directed} if each edge $e \in E$ is an ordered tuple, (i.e. $e = v \to u$ for $v,u \in V$). 
\end{definition}

\begin{definition}
A {\em tournament} $T$ is a directed graph, where for every vertex pair $(u,v) \in V$, there exists a directed edge $u \to v$ or a directed edge $v \to u$ (but not both).
\end{definition}

\begin{definition}
    A \emph{path} of length $k$ in graph $G=(V,E)$ is a sequences of $k$ consecutive edges $(a_1, a_2), (a_2, a_3),..., (a_{k-1}, a_k)$.
\end{definition}

\begin{definition}[King]
We say that a vertex $v$ of a tournament graph is a king if for every other vertex $u$ there is a path from $v$ to $u$ of length at most 2. 
\end{definition}


\begin{definition} [$k$-King Problem2]
The $(k,n)-$ king search problem has as input a tournament graph on $n$ vertices and a number $k \in \mathcal{N}$. A solution to the problem is a set of any $k$ kings in the graph, or $\bot$ if the graph does not contain $k$ kings.
\end{definition}



\begin{notation}
Let $v$ be a vertex. We denote by $N(v)$ the set of all vertices $u$ with an edge $v \to u$. Similarly, we denote by $N^c(v)$ the set of all vertices $u$ with $u \to v$.
\end{notation}

\begin{definition} [\tri]
The \tris inputs a tripartite graph on sets $A,B,C$ with $|A| \in  \Omega(n)$, $|B| \in  \Omega(n)$ and $|C| \in  \Omega(n)$. The goal is to determine if there exists a vertex $a \in A$ such that for all edges $e=(a,c)$ with $a \in A$ and $c \in C$, $e$ belongs to some triangle. 
\end{definition}


\subsubsection*{Query Complexity}
The query model (also known as the Decision Tree model) is one of the simplest computational models. In this model, in the context of tournament graphs, the algorithm makes queries to an oracle, where each query inputs two vertices, $u$ and $v$ and the oracle returns whether $u \to v$ or $v \to u$. The goal is to design an algorithm that makes as few queries as possible. In the randomized setting, the algorithm is allowed to use randomness in order to decide what to query next. Alternatively, a randomized algorithm randomly picks a deterministic decision tree. 

There are many complexity measures of interest in the query model. For instance, the \emph{sensitivity} is the minimum amount of input bits needed to change in order to change the value of the function. The \emph{certificate complexity} is analogous to the non-deterministic complexity -- that is the number of queries an algorithm needs to make to verify the output of a function. There are many other complexity measures, such as block sensitivity, randomized certificate complexity, degree, approximate degree, etc. For total decision problems, all these complexity measures are polynomialy related. 
See, for example, Buhrman and de Wolf~\cite{BuhrmanW02},
Jukna~\cite[Chapter 14]{Juk12} and O'Donnell~\cite[Chapter
8.6]{DBLP:books/daglib/0033652} for a detailed survey of the query model. 

\paragraph{Yao Minmax Principle} When proving lower bounds on the randomized
query complexity of a function $f$ it is often easier to rephrase the problem by
applying Yao's~\cite{Yao77} Minimax Principle. This says that in order to prove a randomized lower bound, it suffices to show an input distribution, and to prove every deterministic algorithm has a lower bound on this distribution.

\subsection{Preliminary Results}
We give a few basic results about kings. Many of these lemmas are used later on.

Many algorithms (upper bounds) of finding kings make use of the \emph{Pivot Lemma}:
\begin{lemma} [Pivot Lemma \cite{AjtaiFHN16}, \cite{BiswasJRS22} \cite{ShenSW03}, \cite{Maurer}, \cite{MandePS23}]\label{lemma:pivot}
For every vertex $u$, it holds that every king in the sub-graph $N^{c}(u)$ is also a king in $G$. We refer to $u$ as a \emph{Pivot}.
\end{lemma}

\begin{proof}
Consider a vertex $v$ that is a king in the subgraph $N^{c}(u)$. By definition, $v$ is within distance 2 to every vertex in $N^c(u)$ and of distance 1 to $u$. Every other vertex in the graph is of distance 1 to $u$ and thus of distance 2 from $v$.
\end{proof}

\begin{lemma}\label{lem:allking} (\cite{Maurer} Thm 14)
There exists a graph where all nodes are kings.
\end{lemma}
\begin{proof}[Sketch] 
Consider the random graph.\footnote{There also exists an explicit construction, see \cite{Maurer} for details.} For a vertex $v$, the probability that $v$ is not a king is the probability that there exists some vertex $u$ with $u \to v$ and $u\to z$ for every $z \in N^c(v)$. For each vertex $u$ this happens with probability $\left(\frac{1}{2}\right)^{|N^c(v)| + 1}$. Due to Chernoff's inequality, with high probability $|N^c(v)| \ge \frac{n}{10}$. Taking the union bound over all vertices $u$, we have that $v$ is a king with high probability. Finally, taking the union bound over all vertices gives the desired result.
    
\end{proof}

\begin{lemma}\cite{Landau53}\cite{Maurer}
A vertex of max outdegree is a king.
\end{lemma}

\begin{proof}
Let $v$ be a vertex of maximum degree. Assume towards a contradiction that $v$ is not a king, that is there exists a vertex $u$ with $u \to v$ and $u \to z$ for each $z \in N(v)$. The degree of $u$ is thus at least $N(v) +1$, a contradiction.
\end{proof}

\begin{lemma} \cite{Maurer}
A tournament has exactly one king, if and only if this king has an outgoing edge to every other vertex.
\end{lemma}

\begin{proof}
    If a vertex, $v$ has an outgoing edge to every other vertex, then no vertex $u$ can reach $v$, and thus $v$ is the only king.

    For the other direction, let $v$ be the only king, and suppose towards a contradiction that $N^c(v) \neq \emptyset$. Finding a king in the subgraph $N^c(v)$ will result in a second king due to \cref{lemma:pivot}, a contradiction.
\end{proof}

\begin{lemma} \cite{Maurer}
\label{lem:2-kings}
No tournament contains exactly 2 kings. 
\end{lemma}

\begin{proof}
Suppose towards a contradiction that a tournament contains two kings, $k_1$ and $k_2$. WLOG, assume that $k_1 \to k_2$. Consider the set $N^c(k_1)$. Due to \cref{lemma:pivot} the king in the subgraph  $N^c(k_1)$ is also a king in $T$. Since $k_1 \to k_2$, $k_2 \notin N^c(k_1)$, the king in $N^c(k_1)$ is a third king. 
\end{proof}

\begin{lemma} \label{lem:1kingdet} \cite{ShenSW03}
There exists a deterministic algorithm for finding a king with $O(n^{3/2})$ queries in the query model.
\end{lemma}

\begin{proof}[Sketch] 
Pick an arbitrary sub-graph of $2\sqrt{n}$ vertices. By a counting argument, in this subgraph there must be a vertex, $v$, of out-degree at least $\sqrt{n}$ (See e.g.\cite{LuWW00}). Pick $v$ as a pivot and consider the set $N^c(v)$. Due to \cref{lemma:pivot} the king in $N^c(v)$ will also be a king in $T$. The algorithm then repeats this process recursively. The running time is $n^{1.5}$: Letting $N$ be the size of the graph at each recursion level, the amount of queries made at each step of the recursion is $N$, and the algorithm eliminates $\sqrt{N}$ vertices in each recursion level.
\end{proof}

\begin{lemma}\cite{ShenSW03}
Every deterministic algorithm in the query model requires $\Omega(n^{4/3})$ queries to find a king.
\end{lemma}
\begin{proof} [Sketch] 

The adversarial strategy is as follows: When an algorithm queries a vertex pair $u,v$, if the outdegree of $u$ is larger than the outdegree of $v$ then direct the edge from $v$ to $u$ and otherwise direct the edge from $u$ to $v$. 

Suppose vertex $k$ is a king. We will calculate the amount of queries that the algorithm must have made. Let $\kout$ be outdegree of $k$. The algorithm must make 
$\displaystyle\sum_{i = 1}^{\kout} i \in \Omega(\kout^2)$ queries, since for every vertex $v$ with an edge directed from $k$ to $v$ there must be at least as many queries made to $v$ as to $k$. 

In order for $k$ to be a king, for each vertex in $ v \in N^c(k)$ there be an edge directed from some $u \in N(k)$. That is a total of:
$ \left(\frac{n - \kout}{\kout} \right)^2 $. Thus the algorithm must make a total queries of
$\kout^2 + \left(\frac{n - \kout}{\kout} \right)^2 $. Minimizing this equation gives the $\Omega(n^{4/3})$ lower bound.
\end{proof}

\begin{theorem} [Theorem \ref{thm:singlekingrand} Restated] 
There exists a (zero-error) randomized algorithm that finds a king in $2n$ queries in expectation.
\end{theorem}

\begin{proof}
We suggest a simple algorithm that works as follows: Pick a vertex $v$ at random. If $N^c(v) = \emptyset$ then return $v$, if not, then repeat on the subgraph $N^c(v)$. Correctness follows immediately from $\cref{lemma:pivot}$.

In order to analyze the expected number of queries, note that in each level of recursion, in expectation, half of the vertices will be eliminated. Formally, we denote as $A_i$ the vertices the algorithm considers at stage $i$ ($A_0 = V$), and note that $|A_i|$ bounds the number of queries performed in stage $i$. Next, by the law of total expectation, we have that \beq \nonumber
\mathbb{E}(A_{i+1}) = \mathbb{E}(\mathbb{E}(A_{i+1} | A_i)) = \mathbb{E}\left(\frac{1}{2}A_i\right)
\eeq

Finally, by the linearity of expectation and induction, we get that the expected number of queries is bounded by
\beq \nonumber
\sum^\infty_{i=0} |A_i| = \sum^\infty_{i=0} \frac{1}{2^i}|V| = 2|V| = 2n
\eeq

\end{proof}

\begin{remark}
Note that no ``sophisticated" data structures are needed to implement the above algorithm efficiently. We simply need to list the nodes of the current subgraph in an array so that it will be possible to pick one at random. Given a vertex $v$  partition the listed vertices so that the first part contains $N^c(v)$ (those pointing into $v$) and continue recursively with the first part of the array (and record the size of  $|N^c(v)|$), as in Quicksort.
\end{remark}
\section{The Complexity of Finding Multiple Kings in the Query Model}

In this section we study the complexity of finding $k$ kings in the query model. We present both upper and lower bounds. 
We generalize the algorithms for finding a single king, and show that the upper bound for finding one king also applies for finding two or three kings. However, the technique breaks down when looking for four or more kings, for which we prove an unconditional lower bound of $\Omega(n^2)$ queries. Formally, we prove \cref{thm:multiplekings} and \cref{thm:ManyKingsLower}. 

\begin{theorem}[Theorem \ref{thm:multiplekings} Restated]
If there exists an algorithm that finds one king in time $T$ then there exists an algorithm that finds $k$ kings, for $k \le 3$ in time $O(T)$.
\end{theorem}
\begin{proof} 
The idea of the algorithm is to find a king, use it as a pivot, and then find a new king in the subgraph. More formally: Find a king $k_1$ in $G$. If there is an edge from $k_1$ to every other vertex, then there can't be an additional king, since every other vertex can't reach $k_1$ in two steps. If this isn't the case, we provide an algorithm for finding two additional kings. First, find a king $k_2$ in $N^c(k_1)$. By \cref{lemma:pivot}, $k_2$ is a king in $G$. Next, consider the subgraph $N^c(k_2)$. Note that this set can't be empty, since if it were empty then $k_1$ wouldn't be a king. Similarly, the king $k_3$ of $N^c(k_2)$ is a king in the entire graph $G$ due to \cref{lemma:pivot}. Note that $k_1 \notin N^c(k_2)$ since by construction $k_2 \in N^c(k_1)$, and thus this third king can't be $k_1$. Further note that the complexity of finding $k_2$ and $k_3$ is at most the complexity of finding $k_1$, since it simply involves finding a king in a smaller subgraph.
\end{proof}

\begin{remark}
We note that the above algorithm cannot be used to find $4$ kings. This is because if we pick $k_3$ as a pivot point, the algorithm may find $k_1$. Indeed, as we see next, the ability to have three non-transitive kings (that is $3$ kings with $k_1 \to k_2 \to k_3 \to k_1)$ is the basis of the $n^2$ lower bound for finding $4$ or more kings. 
\end{remark}


\begin{theorem} [\Cref{thm:ManyKingsLower} Restated]
For every $k\ge 4$ any randomized algorithm for finding $k$ kings has $\Omega(n^2)$ query complexity. 
\end{theorem}

\begin{figure}
\begin{center}
\includegraphics[width=8cm]{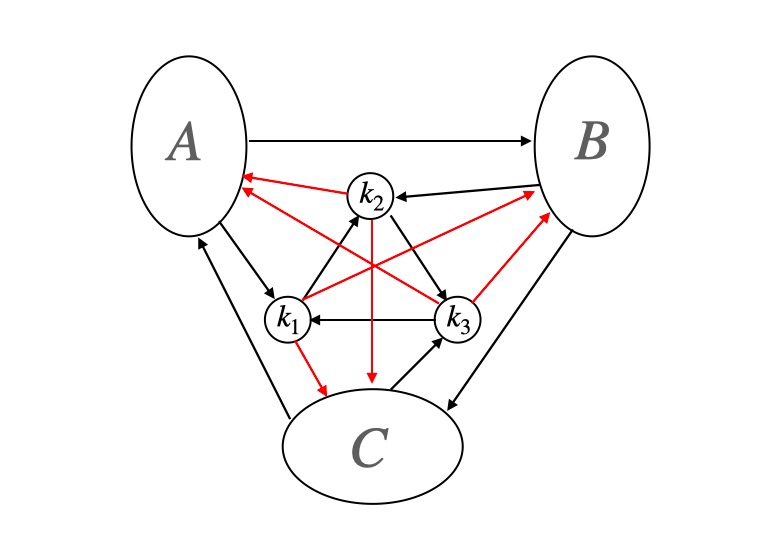}

\caption{If there is no edge between $A,B,C$ to $k_1, k_2, k_3$ then that means that the edges goes from $k_i$ to the set. }\label{fig:n2bound}
\end{center}
\end{figure}

\begin{proof}
We construct a tournament with high sensitivity, that is there are $\Omega(n^2)$ edges where if any is flipped then the number of kings increases. This implies our lower bound, since it suggests the existence of a hard distribution: either the original graph is given or a random edge (among the sensitive ones) is flipped. It is impossible to distinguish these two distributions with less than $\Omega(n^2)$ queries. We can then appeal to the Yao Minmax principle to argue that no query algorithm can succeed in deciding whether there are three or four kings with fewer than $\Omega(n^2)$ queries in expectation. 

Let $k_1, k_2, k_3$ be three vertices satisfying $k_1 \to k_2 \to k_3 \to k_1$. These vertices will each be a king.
The remaining $n-3$ vertices are then grouped into 3 sets each of size $(n-3)/3$. Denote these sets by $A, B, C$. Each set induces an ``all king" subgraph (such a graph exists, see \cref{lem:allking}). For each $a_i \in A$ and $b_i \in B$ direct the edge $a_i \to b_i$. For each $b_i \in B$ and $c_i \in C$ direct the edge $b_i \to c_i$.  For each $c_i \in C$ and $a_i \in A$ direct the edge $c_i \to a_i$. For each $a_i \in A$ $a_i \to k_1$ and $k_2, k_3 \to a_i$. For each $b_i \in B$ $b_i \to k_2$ and $k_1, k_3 \to b_i$. For each $c_i \in C$ $c_i \to k_3$ and $k_1, k_2 \to c_i$. See Figure \ref{fig:n2bound}.

We observe that each $k_i$ is a king and no other vertex is a king. Note that thus far the construction is deterministic.

We call a vetrex $a \in A$ (resp $b\in B$, $c\in C$) a \emph{donkey} if flipping a single edge to any vertex $c \in C$ (resp $a \in A$, $b\in B$) turns $a$ into a king. Indeed, all vertices in $A$,$B$, and $C$ are donkeys: flipping any edge between any vertex $a \in A$ and $c \in C$ results in $a$ becoming a king. Similarly, flipping any edge between $b \in B$ and $a \in A$ results in $b$ becoming a king. Lastly, flipping any edge between $c \in C$ and $b \in B$ results in $c$ becoming a king. 

Consider  $A' \subset A$.  Observe that if we flip any edges between vertices in $A'$ and $C$ then for each $a \in A$ but $a \notin A'$, $a$ is still not a king as it can't reach $k_3$ within two hops. Similarly, vertices in $B$ can't reach $k_1$ in two hops and vertices in $C$ can't reach $k_2$ in two hops

Define $
\Delta$ to be the input distribution where we take this construction and pick uniformly at random $k-3$ vertices. If the chosen vertex is in $A$ (resp $B, C$), then pick a random vertex in $C$ (resp $A$, $B$) and flip the orientation of the edge between these two vertices. 

Consider any deterministic algorithm. Once an algorithm makes $\frac{n}{60}$ queries to some vertex $u$ the algorithm is given for free the orientation all edges adjacent to $u$, and thus as a result the algorithm knows if $u$ is a king. We call a vertex $u$ saturated if the algorithm knows whether $u$ is a king or not, this is done by either the algorithm making $\frac{n}{60}$ queries to it, or by finding a flipped edge in fewer queries. 

Consider an algorithm that has made fewer than $\frac{n^2}{600}$ queries. We will argue that with high probability fewer than $\frac{n}{3}$ vertices will be saturated. Suppose each vertex that has become saturated in fewer than $\frac{n}{60}$ queries has become saturated in 0 queries. We call such a vertex quickly saturated. The probability that a vertex that is a king (Note that a vertex which is not a king can't become saturated in fewer than $\frac{n}{60}$ queries.) becomes saturated in fewer than $\frac{n}{60}$ queries is at most $\frac{1}{10}$, since there are $\frac{n}{3} -1$ edges and each one is equally likely to be flipped. Thus, an algorithm that makes $\frac{n^2}{600}$ queries, will quickly saturate at most $\frac{n}{6000}$ vertices in expectation. The remaining of the saturated vertices each require $\frac{n}{60}$ queries, and thus the algorithm will slowly saturate at most $\frac{n^2/600}{n/60} = \frac{n}{10}$. Thus, in total, the algorithm will saturate in expectation at most $\frac{n}{10} + \frac{n}{6000} < \frac{n}{9}$.

Thus, by Markov's inequality, the probability that the number of saturated vertices is at least $\frac{n}{3}$ is at most $\frac{n/9}{n/3}=\frac{1}{3}$. We conclude that by Yao's Minmax Principle we are done. 
\end{proof}

\paragraph*{Is it possible to get an instance optimal algorithm for finding $k>3$ kings?} An alternative to worst-case complexity is \emph{instance complexity}. Here the goal is to design an algorithm that on each instance performs as well as any algorithm~\cite{GrossmanKN20}. Such algorithms are called \emph{Instance Optimal}. In other words, we would like an algorithm that is competitive with an algorithm that receives as an ``untrusted hint" the full graph, and the number of queries it performs is measured only on correct hints. It turns out that the previous construction also shows that it is impossible to achieve instance optimality. 
\section{The Complexity of Finding Multiple Kings in the RAM Model}

So far, we have seen that finding $k \geq 4$ kings requires $\Omega(n^2)$ queries in the query model, and thus, the trivial algorithm of querying all the edges is optimal. 

Thus, studying the complexity of finding $k \geq 4$ kings is more interesting in the standard Word-RAM model. One can find $k$ kings in $O(n^{\omega})$, which we show is likely to be optimal for large $k$, by making a fine-grained reduction from the \tri. For small $k$, we provide an algorithm running in time $O(n^2k)$.

We now explore the complexity of finding $k \geq 4$ and prove \cref{thm:ManyKingsUpper}.

\subsection{Upper Bounds}
\begin{theorem} [Theorem \ref{thm:ManyKingsUpper} Restated]
There exists a deterministic algorithm for finding $k$ kings that runs in time $$O\left(\min\{n^{\omega}, kn^2\}\right).$$
\end{theorem}

\begin{proof}
We first show the $O(n^{\omega})$ upper bound. Given a tournament $T$, define the matrix, $M$, to be the adjacency matrix of T, that is $M[i,j] = 1$ if $i \to j$ and $M[i,j] = 0$ if $j \to i$. A vertex $i$ is a king iff there is a row in $M^2$ where every entry is non-zero. If $M^2 [i,j] \ge 1$ then that means that there exists some $k$ where $M[i,k] = 1$ and $M[k,j] = 1$. That is, $i \to k \to j$, and thus vertex $i$ can reach $j$ in two hops. Thus, the algorithm simply computes $M^2$ by matrix multiplication and checks whether it has $k$ nonzero rows.

Next, we show the $O(kn^2)$ upper bound. The proof is based on the following lemma:
\begin{lemma} \label{k2}
Let $K$ be a set of vertices ($K$ will be the set of kings found thus far). Let $K^2$ be the set of vertices where each vertex in $K^2$ is of distance at most 2 from every vertex in $K$. A king in the subgraph $K^2$, is also a king in $G$.
\end{lemma}

\begin{proof}
Let $v$ be a king in the subgraph $K^2$. By definition $v$ is of distance at most 2 from every other vertex in $K^2$, and from every other vertex in $K$. Suppose, towards a contradiction, that there exists a vertex $u$ who isn't in $K$ or $K^2$, and $v$ cannot reach $u$ in  one or two hops. Thus it must be that $u \to v$. Furthermore, it must be that $u \to z$ for every $z \in N(v)$. Thus each path of length at most $2$ from $v$ to some $k \in K$ induces a path of the same length from $u$ to $k$, hence $u$ is also in $K^2$, a contradiction to $v$ being a king in $K^2$. 
\end{proof}

The algorithm proceeds in rounds. In each round,  $K$ is the set of kings found thus far, and $K^2$ is the set of vertices $V \setminus K$ such that each vertex in $K^2$ is of distance at most 2 from every vertex in $K$ (i.e\ it can reach all the kings found so far in at most two hops). In each round one vertex $v \in K^2$ is added to $K$, by finding a king in $K^2$ deterministically in time $O(n^{1.5})$ (\cref{lem:1kingdet}). Due to \cref{k2}, $v$ is also a king in the the full tournament. In time $O(n^2)$, $K^2$ is modified to ensure that all vertices in $K^2$ are of distance at most 2 from $v$. This is simply done by checking for every vertex $u$ in $K^2$ if $u$ has an edge to $v$, or if one of the vertices in $N(u)$ has an edge to $v$. Since there are a total of $k$ rounds and in each round $O(n^2)$ time is spent, the total running time is $O(kn^2)$.
\end{proof}



\subsection{Conditional Lower Bound Based On Triangle Detection} \label{ref:TriDetBackground}

Can these upper bounds be improved when $k$ is large?

For polynomial time problems, there are currently no techniques for proving $\Omega(n^{2+\varepsilon})$ lower bounds unconditionally; instead, in \emph{fine-grained complexity}, one seeks conditional lower bounds that are based on the hardness of other, more basic problems.

The most popular assumption for proving lower bounds of the form $\Omega(n^\omega)$ is the conjecture that Triangle detection cannot be solved in $O(n^{\omega-\varepsilon})$ time, for some $\varepsilon>0$ \cite{AbboudW14}. In its most basic form, the problem asks if a given tripartite graph on parts $A,B,C$ contains any triangle $a\in A, b \in B, c \in C$ s.t. $(a,b),(b,c),(a,c)$ are edges. 

The conditional lower bound in this paper is based on a variant of the triangle detection problem in which we change the quantifiers in the problem definition and ask whether there exists a node $a \in A$ that is in a triangle with all its neighbors in $C$.


\begin{definition} [\tri]
The $(n,|A|)-$\tris inputs a tripartite graph on sets $A,B,C$ with $|A| \in  \Omega(n)$, $|B| \in  \Omega(n)$ and $|C| \in  \Omega(n)$. The goal is to determine if there exists a vertex $a \in A$ such that for all edges $e=(a,c)$ with $c \in C$, $e$ belongs to some triangle. 
\end{definition}

Throughout our paper we will be using the following definition, which due to basic fine grained reduction techniques, is equivalent:

\begin{definition}
The $(n,|A|)-$\tris inputs a tripartite graph on sets $A,B,C$ with  $|A| \in  \Omega(n)$, $|B| \in  \Omega(n)$ and $|C| \in  \Omega(n)$. Furthermore, each vertex has degree at least 1 to each of its neighboring sets. The goal is to determine if there exists a vertex $a \in A$ such that for all edges $e=(a,c)$ with $c \in C$, $e$ belongs to some triangle. 
\end{definition}

While there is no reduction from the basic Triangle detection problem to the \tri, it is natural to conjecture they are equally hard.
None of the currently known existing techniques is able to benefit from the difference in quantifiers.

The concept of assuming that the core problems of fine-grained complexity remain hard when changing their quantifiers was suggested in \cite{AVW16}, and has since been employed multiple times \cite{BC20,ABHS22,HuangLSW23,ADLW23} when reducing to problems that encode certain quantifications (as is the case in the current paper).  Further justification for this practice was provided in \cite{CarmosinoGIMPS16}, where it was argued that the hardness of such problems cannot be based on the hardness basic forms of the problems due to differences in their nondeterministic time complexity.
We are not aware of a prior work that uses the specific variant in our \tri.

\begin{theorem}[Theorem \ref{thm:finegrained} Restated]
There is a linear time fine-grained reduction between $(n, |A|)-$ \tris to the (k,n')-king problem, for $n' = n + |A| +2$ and $k = n - |A| +3$.
\end{theorem}

\begin{proof}
Given a tri-partite graph on set of vertices $A_G$, $B_G$, $C_G$, with edge set $E_G$, define the following tournament graph on $n'$ vertices:

Let $A_T, A_T', B_T, C_T$ each be a set of vertices, with $|A_T| = |A_T'| = |A_G|$. And $|B_T| = |B_G|$ and $|C_T| = |C_G|$. Let $x_2 \to x_1$ be two additional vertices. 

Next, we define the orientation of the edges:

Within each $A_T$, $A_T'$, $B_T$, and $C_T$ each of the edges induces an all-kings sub graph (Such a graph exists, see \cref{lem:allking}). For relationships between different sets define: \begin{itemize}
\item For all $a_T \in A_T$, $b_T\in B_T$ orient the edge $a_T \to b_T$ iff $(a,b) \in E$. 
\item For all $b_T \in B_T$, $C_T\in C$ orient the edge $b_T \to c_T$ iff $(b,c) \in E$.  
\item For all $c_T \in C$, $a_T'\in A'$ orient the edge $c_T \to a_T'$ iff $(c,a) \in E$. 
\item For all $a_i \in A_T$, $a_j'\in A_T'$ orient the edge $a \to a'$ iff $i =j$.
\item For all $c_T \in C_T$, $a_T\in A_T$ orient the edge $c_T \to a_T$. 
\item For all $b_T \in B_T$, $a_T'\in B_T$ orient the edge $b_T \to a_T'$. 
\item For all $v \in A_T'\cup B_T $, orient the edge $x_1 \to v$.
\item For all $v \in A_T \cup C_T \cup A_T' $ orient the edge $x_2 \to v$.
\item For all $b_T \in B_T$, orient the edge $b_T \to x_2$.
\end{itemize}

\begin{figure}
\begin{center}
\includegraphics[width=6cm]{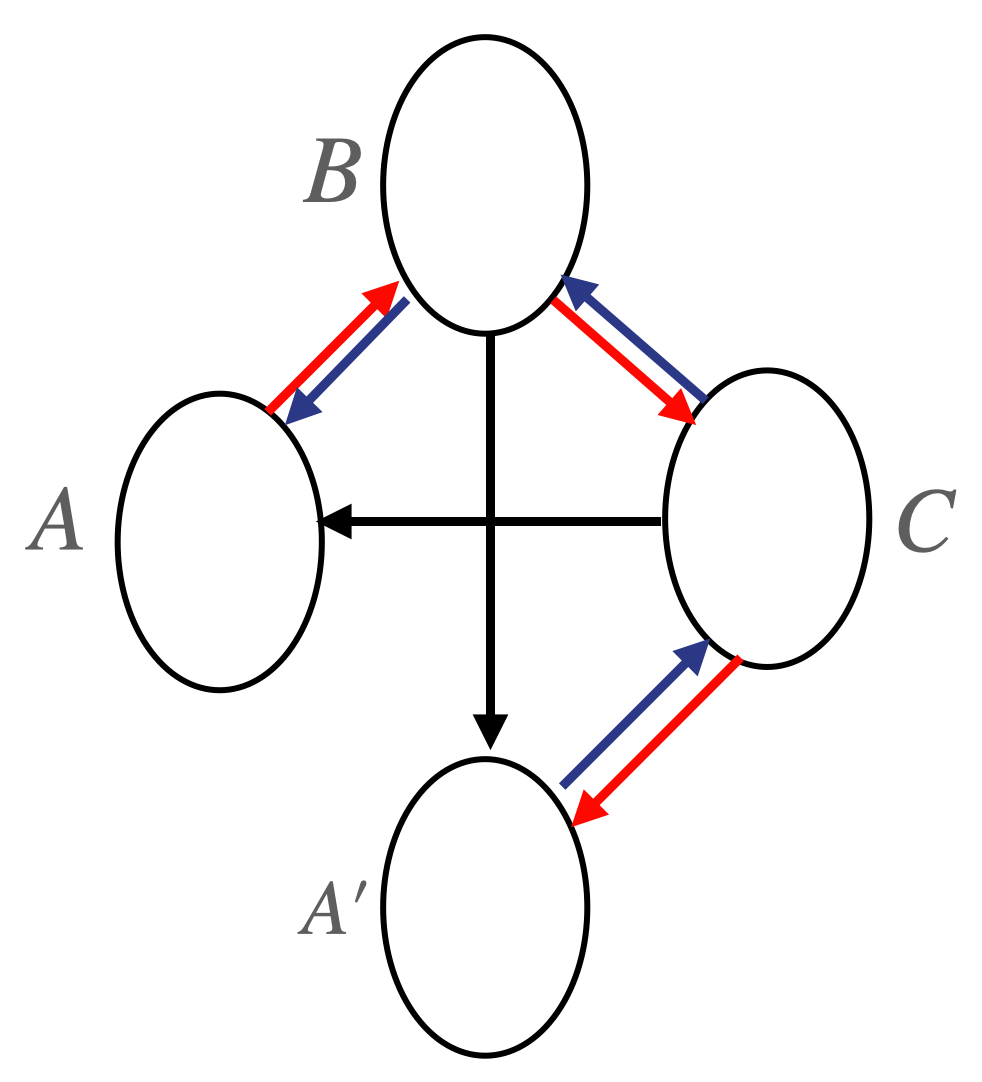}

\caption{Relationship between $A$, $A'$ $B$ and $C$. A  red arrow indicates that this is the direction of the edge if the two vertices have an edge in $G$. A blue arrow indicates that this is the direction of the edge if there is no edge in $G$. A black arrow indicates that this is the direction of the edge regardless of $G$. For $A$ and $A'$, we have $a_i \to a'_j$ if $i = j$ and otherwise $a'_j \to a_i$.\\}\label{fig:FineGrained}
\end{center}
\end{figure}
    
\begin{figure}
\begin{center}
\includegraphics[width=6cm]{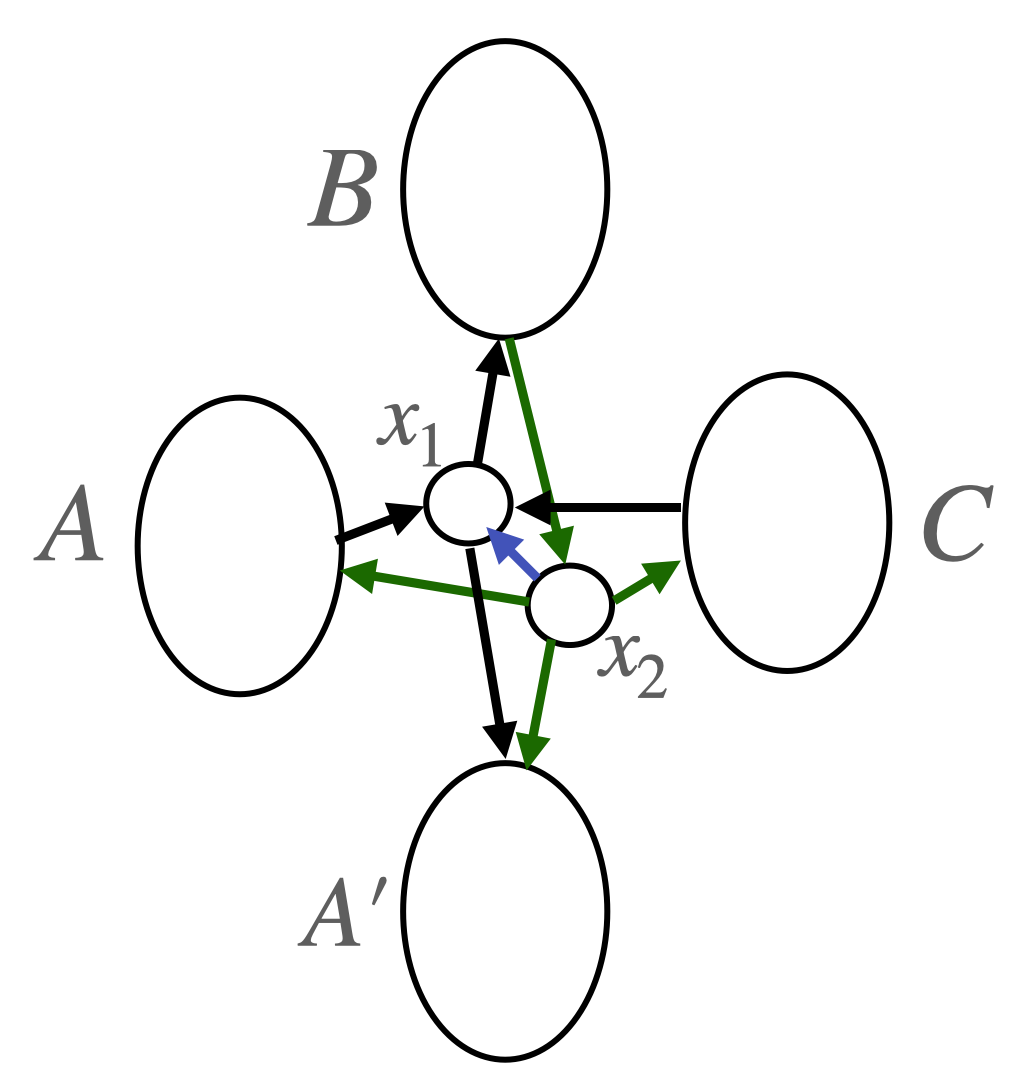}

\caption{Relationship between $x_1$, $x_2$ and the remaining sets.\\}
\label{fig:FineGrained2}
\end{center}
\end{figure}

See Figure \ref{fig:FineGrained} and Figure \ref{fig:FineGrained2}. 

\begin{claim}
 All vertices $b_T \in B_T$ and $c_T \in C_T$ are kings, and also $x_1$ and $x_2$ are kings.
\end{claim}
\begin{proof}
Let $b_T \in B_T$. It reaches $x_2$ directly, and from there it reaches $C,A, x_1$ in a single step. In addition, it reaches each other node in $B$ within two steps with the edges on the induced subgraph on $B$, which is the all-kings subgraph by construction.

Let $c_T \in C_T$. $c_T$ can reach every vertex $b,a' \in B,A'$ in two steps via $x_1$. $c$ reaches each $a \in A$ and $x_1$ in one hop. $c$ reaches $x_2$ in two steps via some vertex in $B$ (each vertex in $c$ is connected to some vertex in $B$).

\end{proof}

\begin{observation}
No vertex $a_T' \in A'_T$ is a king. 
\end{observation}

\begin{proof}
Observe that $a_T'$ can't reach $x_2$ in two steps. 
\end{proof}

\begin{claim}
A vertex $a_T \in A_T$ is a king if and only if for all $c_G \in C_G$, with $e = (a,c) \in E$, $e$ is part of a triangle in $G$.
\end{claim}

\begin{proof}
First we will show that $a_T$ is within distance two from every vertex that isn't in $C_T$. Note that $a$ can reach each vertex $b_T \in B_T$ and $a'_T \in A'_T$ in two steps via the vertex $x_1$. Also, $a_T$ can reach $x_2$ in two steps via some vertex $b_T \in B_T$, as by definition $a_G$ has at least one neighbor in $B_G$, and each vertex $b_T \in B_T$ has an edge to $x_2$.

Let $a_G \in A_G$ such that for every $e = (a_G,c_G)$, the edge $e$ is in a triangle. We will show that the corresponding vertex $a_T$ is king.

Let $c_T \in C_T$ such that for the corresponding vertex $c \in C_G$ there is an $e = (a_G,c_G) \in E_G$. Then, since this edge is on a triangle, there are edges $(a_G,b_G), (b_G,c_G) \in E_G$, which imply corresponding edges in $E_T$, which in turn imply a path of length $2$ in $T$ from $a_T$ to $c_T$.

Let $c_T \in C_T$ such that for the corresponding vertex $c \in C_G$ there isn't an $e = (a_G,c_G) \in E_G$. Let $a'_T \in A'$ be the mirror vertex of $a_T$. In this case, $a_T$ can reach $c_T$ via $a'_T$. This holds as by definition there is an edge from $a_T$ to $a'_T$, and since $(a_G,c_G) \not\in E_G$, $(a'_T,c_T) \in E_T$.

Suppose $a$ is a king. Let $e = (a,c) \in E_G$, we want to show that it is part of a triangle.

Since $a_T$ is a king, it must be within distance 2 from $c_T$. Since there is no edge from $x_1$ to $c_T$, and no edge from $a_T$ to $x_2$, it must be the case that $a_T$ reaches $c_T$ in distance 2 via $A_T'$ or $B_T$. Since $e \in E$, $a_T$ can't reach $c_T$ through $A'$. This is because by construction the only vertex $a_T' \in A_T'$ accessible from $a_T$ doesn't have an edge to $c_T$. Thus $a_T$ reaches $c_T$ via some vertex $b_T \in B_T$, implying that $(a_G,b_G), (b_G,c_G), e$ is a triangle. 




\end{proof}
Thus we conclude that a vertex $a \in A_T$ is a king precisely when the corresponding vertex $a \in G$ satisfies the requirement of the \tri: that each edge $(a,c)$ belongs to some triangle. 
\end{proof}

\begin{remark} Our reduction not only shows that the $O(n^{\omega})$ bound is conditionally tight for large $k$, but also that ``combinatorial'' techniques (as employed in our $kn^2$ algorithm) are unlikely to achieve $n^{3-\varepsilon}$ time. This is because such running times are conjectured to be impossible for Triangle detection \cite{WilliamsW18,AbboudW14}, and there is no reason to suppose that the \tris is easier. We refer to \cite{AbboudFKLM23} for more background on combinatorial algorithms.

\end{remark}

\bibliographystyle{alpha}
\bibliography{main}


\end{document}